# Design and advancement status of the Beam Expander Testing X-ray facility (BEaTriX)

D. Spiga[1§], C. Pelliciari[1], B. Salmaso[1], L. Arcangeli[2], G. Bianucci[2], C. Ferrari[3], M. Ghigo[1], G. Pareschi[1], M. Rossi[2], G. Tagliaferri[1], G. Valsecchi[2], G. Vecchi[1], A. Zappettini[3]

[1]INAF / Brera Astronomical Observatory, Via Bianchi 46, 23807 Merate (Italy)
[2]Media Lario S.r.l., loc. Pascolo, 23842 Bosisio Parini (LC)
[3]CNR-IMEM, Parco Area delle Scienze 37/A, 43124 Parma (Italy)

**ABSTRACT**

The BEaTriX (Beam Expander Testing X-ray facility) project is an X-ray apparatus under construction at INAF/OAB to generate a broad (200×60 mm$^2$), uniform and low-divergent X-ray beam within a small lab (6×15 m$^2$). BEaTriX will consist of an X-ray source in the focus a grazing incidence paraboloidal mirror to obtain a parallel beam, followed by a crystal monochromation system and by an asymmetrically-cut diffracting crystal to perform the beam expansion to the desired size. Once completed, BEaTriX will be used to directly perform the quality control of focusing modules of large X-ray optics such as those for the ATHENA X-ray observatory, based on either Silicon Pore Optics (baseline) or Slumped Glass Optics (alternative), and will thereby enable a direct quality control of angular resolution and effective area on a number of mirror modules in a short time, in full X-ray illumination and without being affected by the finite distance of the X-ray source. However, since the individual mirror modules for ATHENA will have an optical quality of 3-4 arcsec HEW or better, BEaTriX is required to produce a broad beam with divergence below 1-2 arcsec, and sufficient flux to quickly characterize the PSF of the module without being significantly affected by statistical uncertainties. Therefore, the optical components of BEaTriX have to be selected and/or manufactured with excellent optical properties in order to guarantee the final performance of the system. In this paper we report the final design of the facility and a detailed performance simulation.

## 1. INTRODUCTION

Parallel and wide X-ray beams are required for several applications: wide-aperture X-ray optics, diffractive lenses, medical imaging, and X-ray tomography to name a few. As of today, broad and quasi-parallel beams can be obtained using distant, powerful, and isotropic sources. However, large X-ray facilities operating in high vacuum are required - such as MPE/PANTER,[1] or the longest beamlines at the largest synchrotron light sources such as BL20B2 at SPring-8[2] - to efficiently test the entire (or a significant fraction of) aperture in X-ray optical systems. This can pose a challenge for X-ray tests of optical assemblies based on a modular approach, as the ATHENA X-ray telescope.[3] ATHENA's optical system will have a 3 m outer diameter, a 12 m focal length, a 2 m$^2$ effective area at 1 keV, an angular resolution better than 5 arcsec half-energy width (HEW) at 1 keV, and better than 10 arcsec at 6 keV. In order to reach these performances, more than 1000 X-ray Optical Units (XOUs) based on Silicon Pore Optics[4] have to be manufactured, tested, and qualified before integration into the supporting structure. Each pore in each XOU reproduces the Wolter-I (parabola+hyperbola) configuration and, owing to the wedged ribs, all the pores in an XOU converge to a single focus. Each XOU integrated into ATHENA has to be accurately co-axially and co-focally aligned, but some alignment error is unavoidable. Hence, each individual, flight-grade XOU must have an angular resolution better than 5 arcsec, and only the most performing XOUs out of the industrial production line can be selected to be integrated. Indeed, because of the small pore size (a few mm$^2$), metrological tests on the XOU stacks are extremely difficult, and only an optical performance measurement in X-rays can represent a reliable qualification test. This also offers the possibility to reliably align the parabolic and the hyperbolic stack of the XOU, as the maximization of the effective area - corresponding to the best align condition - can be checked at-wavelength. Tests is UV light would enable this alignment, but would also be heavily affected by aperture diffraction by the rib/membrane obstruction,[5] making it impossible to measure the HEW expected in X-rays.

---

[§] contact author: daniele.spiga@brera.inaf.it

To date, X-ray tests of SPO XOUs are performed in pencil beam setup at the PTB lab of the BESSY synchrotron light source,[6] but the point spread function (PSF) measurement requires reconstruction because only a pore at a time can be illuminated, and the complete scan of a XOU aperture is time consuming. A measurement at PANTER would enable a true full-illumination test, possibly compensating the beam divergence via a diffractive X-ray lens.[7] However, PANTER is a very large facility and cannot be employed to routinely perform the functional tests of more than 1000 XOUs (plus the discarded and spare ones). Finally, both for BESSY and PANTER, the testing facility cannot be moved near the XOU production site.

We have therefore designed[8] a compact facility named BEaTriX[9] (*Beam Expander Testing X-ray facility*) to generate a broad, parallel, uniform, monochromatic, and polarized X-ray beam. The main science driver is testing ATHENA XOUs in full illumination, with reduced costs and in a time short enough to sustain a production rate of a few XOU modules per day, and so return a prompt feedback to the manufacturer. BEaTriX is being constructed at INAF/OAB, to be possibly replicated at the industrial production site of the XOUs. Specifications include:

i) an X-ray beam wider than 90 mm and higher than 55 mm, to fully illuminate the largest apertures of XOUs;
ii) very low residual divergence (< 1.5 arcsec HEW), to reliably characterize XOU PSFs with an intrinsic HEW of 3-4 arcsec. The collimation in the vertical plane is especially important because it determines the beam spread in the XOU incidence plane, in which the elongating effect of profile/surface defects on the PSF is mostly seen;
iii) good uniformity of the beam, to equally characterize the entire optical surface;
iv) highly monochromatic energy, selectable between 1.49 keV and 4.51 keV;
v) compact size to fit a small laboratory (6 m × 15 m), and reduce the test chamber evacuation time.

In this paper we review an updated design[10] of BEaTriX (Sect. 2) and show the advancement status of component fabrication (Sect. 3). We also show an optical simulation of the system with the reviewed design (Sect. 4).

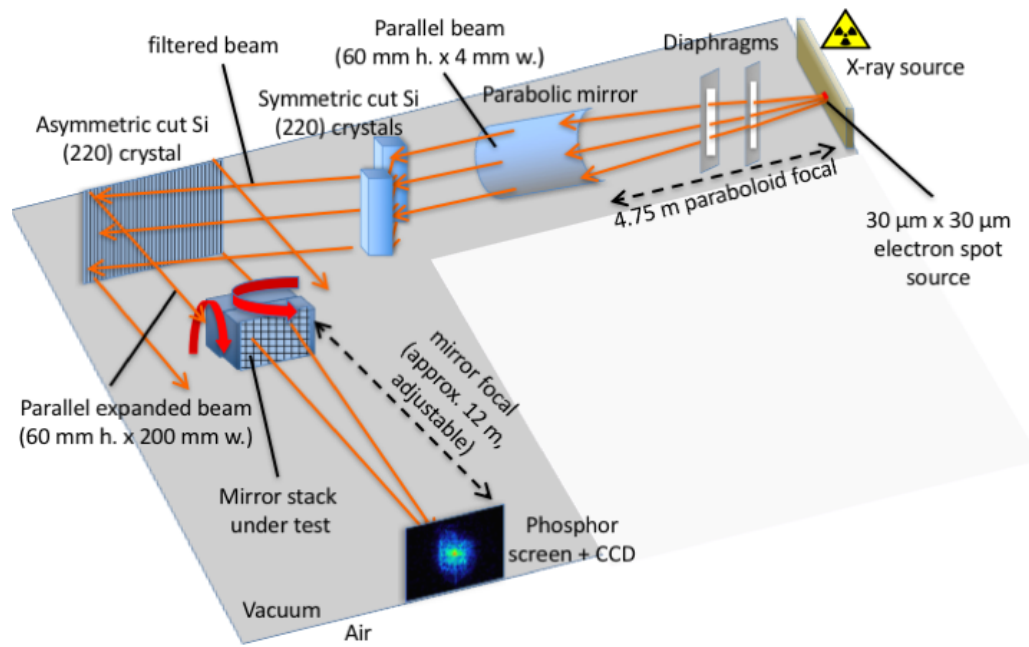

Fig. 1: schematic design of the BEaTriX facility in the 4.51 keV setup. The beam diverging from a microfocus X-ray source is collimated by a precisely figured paraboloidal mirror (Fig. 2) with the source in its focus. After reflection, the beam has a 4 mm (width) x 60 mm (height) size and the subsequent diffraction operated by a pair of symmetric silicon crystals isolates the 4.51 keV fluorescence line (Fig. 3). The horizontal expansion of the beam is done by another silicon crystal, asymmetrically cut with respect to the (220) planes (Fig. 4). The expanded beam (200 mm × 60 mm) can be used to illuminate the aperture of the focusing device under test, and the focus is directly observed by means of an imaging detector.

## 2. PROJECT DESIGN REVIEW

### 2.1. Optical component description

Once completed, BEaTriX will exploit a combination of grazing-incidence mirrors and symmetric/asymmetric diffracting crystals to expand an X-ray beam to the required size. The detailed facility layout, shown in Fig. 1, includes:

1) *A microfocus X-ray source* with either aluminum (1.49 keV setup) or titanium (4.51 keV setup) anode to generate an intense X-ray beam including the fluorescence lines and a bremsstrahlung continuum: the electron spot source has to be extremely small (30 μm x 30 μm) to ensure the required vertical collimation. *Microfocus* sources of this kind are commercially available: some models generate a copious X-ray flux (on the order of $10^{10}$ - $10^{11}$ ph/sec/sterad) with a very low power consumption (a few tens of watts).

2) *A paraboloidal, grazing incidence mirror* with a 4.75 m focal length (measured from the nearest side to source) and a 6 cm-high illuminated surface (Fig. 2, left). The X-ray source is located just in the focus of the paraboloid within a ± 1 mm tolerance along the axis and a 50 μm tolerance in the lateral displacement.[10] The mirror is to be coated with a platinum layer (30 nm) and an amorphous carbon overcoating (3 nm) to enhance the low-energy reflectivity.[11] Forming with rays a grazing incidence angle of 0.9 deg, the mirror returns high reflectivity up to 6.5 keV (Fig. 2, right). After reflection, the beam becomes collimated, parallel, with a cross-section in the shape of a crescent, in which a 4 mm wide and 60 mm high rectangle can be inscribed.[10] This rectangular beam, after monochromation, finally impinges on the asymmetric crystal (Fig. 4) and is eventually expanded to the 200 mm × 60 mm size (Fig. 15). Details on the ongoing fabrication of the paraboloidal mirror are reported in Sect. 3.1.

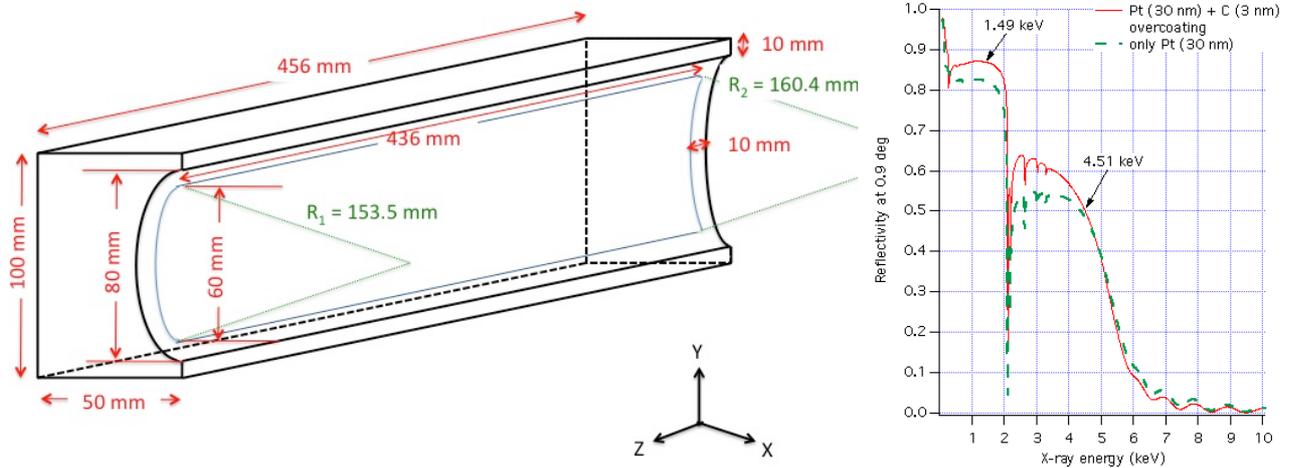

Fig. 2: (left) a drawing of the paraboloidal mirror. The focus is located 4750 mm from the mirror end with smaller curvature radius, in the direction of the positive *z* axis. (right) calculated mirror reflectivity.

3) *A pair (to be possibly increased to two pairs) of symmetrically-cut crystals* (i.e., with diffraction planes parallel to the surface) to filter the Kα line in the X-ray source, but with no change in the beam size. For a high collimation of the X-ray beam, excellent (< 0.5 eV FWHM) monochromation is essential. At 4.51 keV, monocrystalline silicon (220) can be used (1.91 Å d-spacing, corresponding to a 45.85 deg Bragg angle) because the rocking curve shows a sharp peak (Fig. 3, red line) reaching a 95% reflectivity. The collimating mirror acts as a low-pass energy filter and so rules out higher harmonics like (440). The monochromation level can be further improved, employing two diffractions in sequence with crystals in parallel interfaces: in this way, the peak region has still high reflectivity, but the tails are suppressed. The combined reflectivity (Fig. 3, black line) is higher than 80% in a 3 arcsec-wide angular range around the peak. This means that, in order to have high diffraction efficiency, the paraboloidal mirror has to spread the rays in the incidence plane by less than $\Delta\theta_B$ = 3 arcsec: the energy filtering degradation corresponding to this angular acceptance is < 0.1 eV, smaller than the natural X-ray line width and not significantly affecting the final beam collimation. These considerations entail a 3 arcsec HEW required quality for the collimating mirror, which shall therefore be figured and polished in compliance to this requirement.

Another advantage offered by this configuration is that the final beam direction remains parallel to the initial one and no folded vacuum tube is required to follow the beam at this point. Finally, we note that the incidence angle is very near the polarization angle, so the beam is not only filtered spectrally, but also strongly *polarized* (97% after the first diffraction) in the s-plane, i.e., the vertical plane of the facility. While this is not expected to affect the PSF characterization of grazing-incidence mirrors (which reflectivity is fairly independent of the polarization state), it can be useful to perform X-ray polarimetry experiments.[12]

Silicon cannot efficiently diffract 1.49 keV X-rays. For the setup corresponding to this energy, it should be replaced by ADP (101) - Ammonium Dihydrogen Phosphate, $NH_4H_2PO_4$ (5.32 Å d-spacing, corresponding to a 51.46 deg Bragg angle). Also in this case, there is a sensitive polarization effect, although less marked (74%). The higher harmonics in the ADP crystal will be avoided tilting the first crystal by a small angle, in order to slightly shift the Bragg peaks at 1.49 keV, 2.98 keV, …. As the rocking curves for the (202), (303), (404) order are much narrower than (101), only the lowest order survives after two consecutive diffractions.

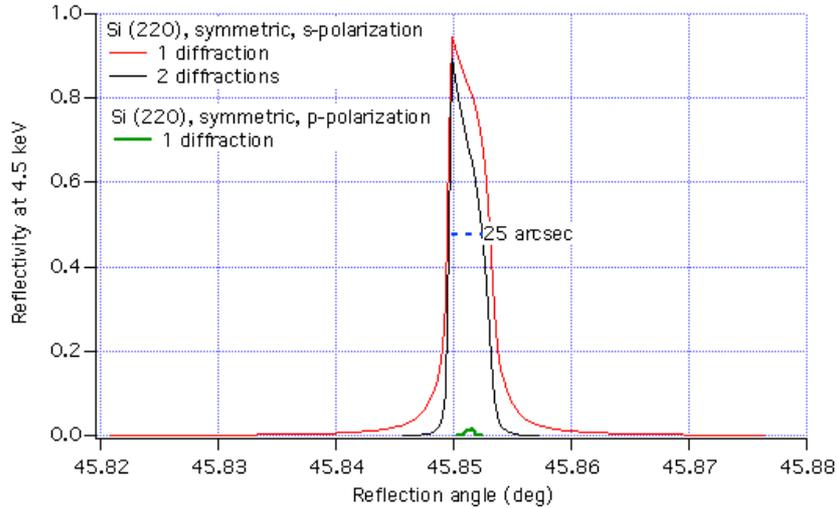

Fig. 3: the (220) rocking curve of the symmetric silicon crystal at 4.51 keV, in single and double diffraction. The incidence angle is very close to the polarization angle and the beam becomes polarized in the tangential plane. Two consecutive diffractions suppress the rocking curve wings and so improve the monochromation considerably.

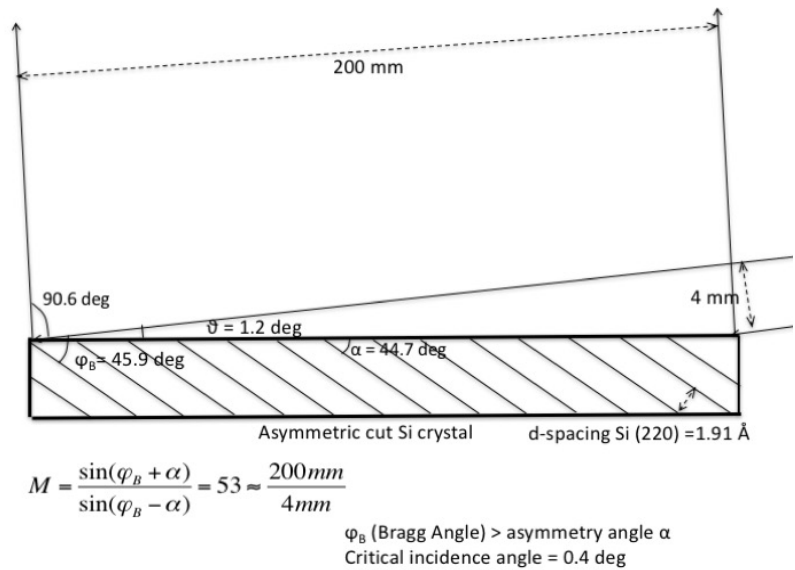

Fig. 4: principle of X-ray beam expansion via an asymmetrically-cut crystal. The incidence angle is larger than the critical incidence angle at 4.51 keV (0.4 deg) on silicon to avoid total external reflection.

4) *An asymmetrically-cut crystal* (i.e., with diffraction planes tilted with respect to the surface)[13] to expand the monochromatic and collimated beam in the horizontal direction. In the 4.51 keV setup, shown in Fig. 4, the silicon crystal has the (220) diffraction planes tilted by 44.7 deg with respect to the outer surface. The beam, with an effective width of 4 mm, impinges in grazing incidence on the crystal surface, but the incidence angle is larger than the critical incidence angle of silicon at 4.51 keV; hence, the total external reflection is avoided. The 60 mm × 4 mm beam is spread over the crystal and diffracted with respect to the tilted (220) planes: because of the incidence angle close to 45 deg on the Bragg planes, the beam is reflected near a right angle off-surface, and the horizontal beam size is expanded by a factor of ~50. The final beam has thereby a 60 mm (v.) × 200 mm (h.) size, fulfilling the beam size request, and can be used to fully illuminate the aperture of the XOU under test, with a vertical divergence of 1.3 arcsec (determined by the source size and the mirror focal length) and a horizontal collimation < 1.5 arcsec HEW (determined by the width of the exit rocking curve of the asymmetric crystal). This is shown in Fig. 5, where the reflectivity curve is plotted as a function of the entrance (left) and of the exit (right) angle. In asymmetric geometry, the exit rocking curve width is 50 times smaller by than the one in entrance; therefore, even if the beam is imperfectly collimated at the exit of the monochromator, the final collimation degree cannot be worse than the rocking curve width in Fig. 5, right. In practice, the final collimation will be expectedly better than 1.5 arcsec, because only the central part (~3 arcsec wide) of the rocking curve in Fig. 5, left, will be used. Details on the ongoing fabrication of the asymmetric crystal are given in Sect. 3.2.

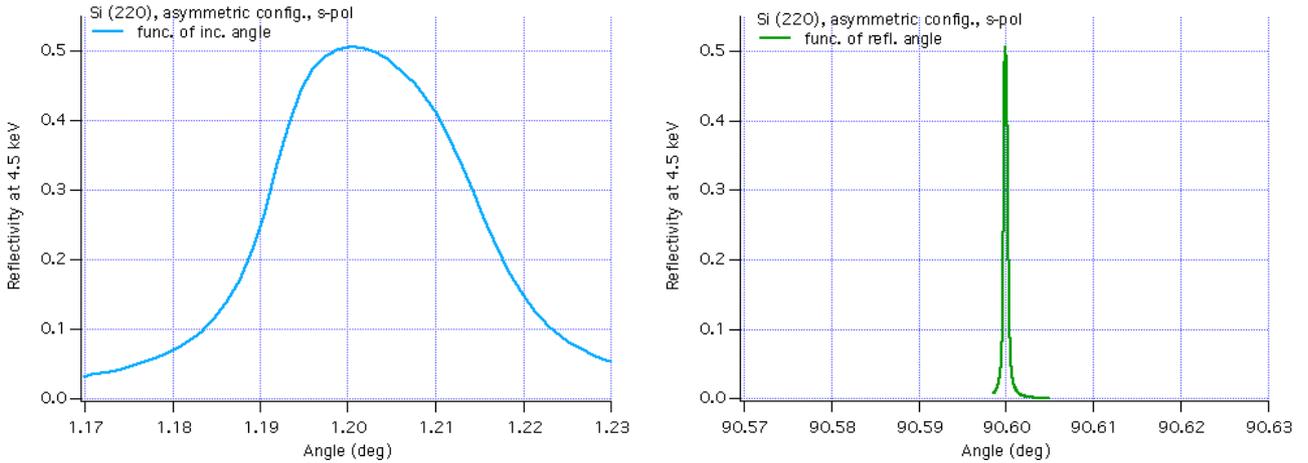

Fig. 5: diffracted intensity by the silicon crystal (220) at 4.51 keV in asymmetric configuration (Fig. 4), as a function of the incidence angle (left) and as a function of the reflection angle (right). The asymmetric diffraction improves the horizontal collimation of the beam to a 2 arcsec FWHM (approx. 1.5 arcsec HEW). The diffracted intensity for the p-polarization is 500 times smaller and is not shown.

5) *A sample alignment stage* to support and precisely align in the expanded X-ray beam the XOU under test. The alignment stage needs two translation motors to move the XOU in the transverse beam section to 1 μm accuracy, a translation motor along the beam to adjust the best focus on the detector, and three rotation stages (< 1 arcsec accurate) to maximize the doubly-reflected beam intensity, a condition corresponding to the best XOU alignment. A more complex system will be obviously required to perform the alignment of the parabolic to the hyperbolic segment of the XOU. Clearly, also the paraboloidal mirror and the diffracting crystals require appropriate alignment stages.

6) *An imaging detector* in the focus of the XOU to directly measure its PSF. To avoid the complication of keeping a camera under vacuum and ease its movement in all directions, the simplest solution is represented by a phosphor window and a CCD sensitive to phosphorescent light to record the image. The camera should have a 10 μm spatial resolution (corresponding to 0.2 arcsec) in order to oversample a few arcsec wide PSF. The field of the camera should cover approximately 10 arcmin (a 34 mm-sided field at a 12 m distance).

## 2.2. Outer facility design

Viewed from outside (Fig. 6), BEaTriX will be composed by two arms, made of vacuum tubes and chambers, roughly forming an "L" shape and enclosing the optical components listed in Sect. 2.3. The corner corresponds to the asymmetric crystal where the beam is approximately deviated by 90 deg in the horizontal plane. The short arm, approx. 6 m long and parallel to the laboratory basement, is responsible for the beam expansion and encloses the source and the optical components. The long arm (~13 m) will include the XOU handling stage, followed by a 12 m-long tube that propagates X-rays focused by the XOU to the focal plane. In order to follow the focused beam - that will be deflected downwards by the double reflection on the XOU - the 12 m long tube shall be tilted in the vertical plane in a 0 - 7.2 deg range, corresponding to the direct beam detection and to the angular deviation by the largest radii of ATHENA. The short arm, in contrast, will be steered horizontally to adjust the angle between the two arms when the testing energy is changed, and so satisfy the Bragg law with the crystals in use. Some vacuum joints will be flexible to enable the angle setting. All the pipeline will be mounted in a self-bearing structure, with the vacuum tubes counterbalanced with calibrated weights in order to minimize the motor loads. The tubes shall be built modular to enable intra-focal measurements.

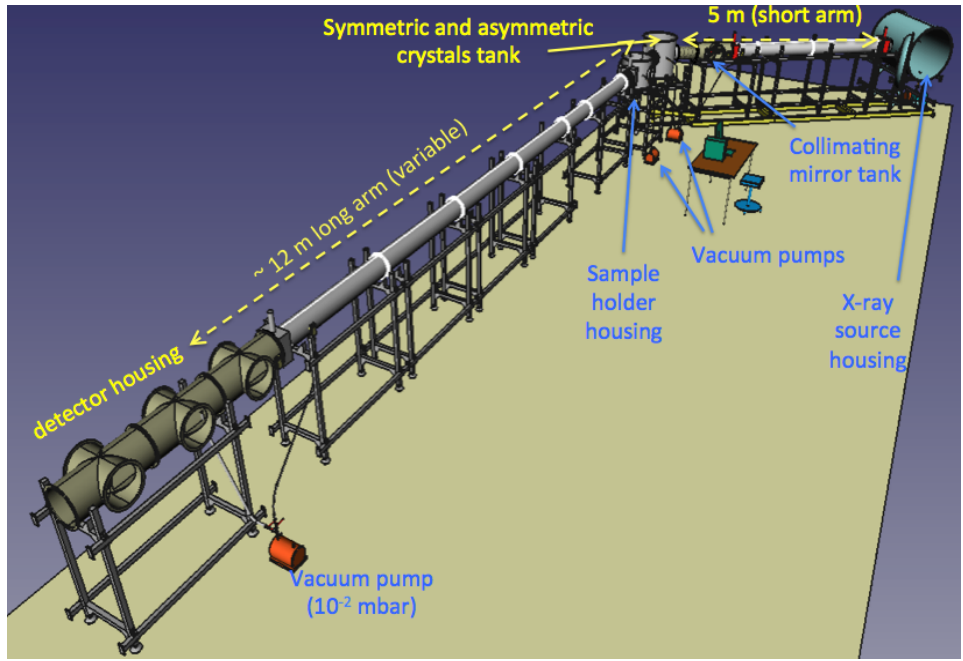

Fig. 6: outer (simplified) view of the BEaTriX facility. The optical elements needed to generate (X-ray source), filter, polarize and expand the beam (parabolic mirror, collimators, symmetric crystals, asymmetric crystal) are all enclosed in the 5 m vacuum tube in the background and in the central tank ("short arm"). The 12 m vacuum tube (the "long arm") allows the rays to be focused on the detector. The short arm can be steered in the horizontal plane to fit the Bragg angle on the asymmetric crystal, while the long arm can be tilted vertically to fit the reflection angle of the XOU under test.

Finally, all the system needs to be evacuated to avoid X-ray absorption in air. The total range of X-rays from source to detector at the end of the 12 m long tube is 17 m. Fortunately, the reduced dimensions of the facility require only a moderate vacuum level: a previous simulation[9] showed that even a 0.1 mbar residual pressure would suffice to preserve 75% of the beam intensity at 1.49 keV and more than 98% at 4.51 keV. In this design, four dry pumps can be used to reach a vacuum level of $10^{-2}$ mbar and independently evacuate four sections of the facility: of these, only the XOU vacuum tank will periodically be vented/evacuated at the operation time, with a considerable time saving when the XOU under test is to be changed. All the system will be mounted in a ground-based laboratory at 20 °C temperature, and stable within 1 °C. A clean tent (ISO4 class or better) mounted above the sample holder housing will allow us mounting/removing the XOU under test without dust contaminations.

## 3. ADVANCEMENT STATUS

### 3.1. The collimating mirror

As detailed in Sect. 2, the divergent X-ray beam is made parallel and expanded in the vertical direction via a grazing reflection on a parabolic mirror. The figuring accuracy of the mirror is essential to guarantee high efficiency in the subsequent spectral filtering by the symmetric crystals. For this reason, the mirror needs to be highly polished and figured aiming at a maximum tolerable HEW of 3 arcsec, and to be thermally and mechanically stable during the alignment and the operation. The material selected is fused quartz HOQ 310, a material easy to polish and figure to an excellent level, characterized by an extremely low level of impurities and inclusions, and very low coefficient of thermal expansion ($0.5\times10^{-6}$ K$^{-1}$).

The first step in the paraboloidal mirror development has been the determination of manufacturing tolerances: doing this, we are primarily interested in the longitudinal profile errors, because in grazing incidence the roundness errors have a weight two orders of magnitude smaller in the PSF broadening. We have therefore created a number of simulated profile errors superimposing elementary harmonic components with spatial frequencies $k/L$, where $L = 436$ mm and $k = 1, 2, ...$ down to a spatial wavelength of 2 mm. The harmonic amplitudes are in proportion to the spatial periods, varying the relative phase of the components at random several times. The resulting profiles have been superimposed to rough profiles constructed from a PSD (power spectral density) in the shape of a power-law $P(f) = K_n/f^n$ in the bandwidth 1 mm – 10 μm with variable $n$ and $K_n$ parameters that have been varied – together with the amplitude/period ratio of the figure error components - until a slope value returning the HEW < 3 arcsec requirement at $E < 4.51$ keV was found. Doing this, the PSF was computed from the modeled profiles using the WISE code[14] to self-consistently compute the PSF from figure error and roughness. The results are shown in Fig. 7; the tolerable value for the amplitude/period ratio of harmonic components is 1 μrad, and the roughness PSD shall be below the power law-model with $n = 1.05$ and $K_n = 18$ nm$^3$μm$^{-1.05}$. This means a final surface quality $\sigma_m < 3$ μrad slope rms at spatial wavelengths larger than 2 mm, and a roughness rms $\sigma < 2$ Å at spatial periods smaller than 1 mm. One of the infinitely possible profile errors within the tolerances is shown in Fig. 8.

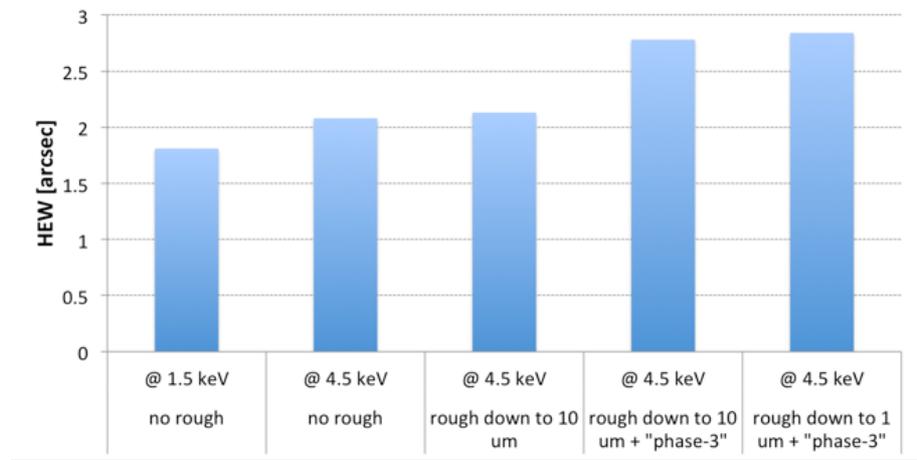

Fig. 7: mirror HEW expected from a proper combination of sinusoidal errors with spatial periods $L$ (in mm) from the mirror length down to 2 mm and amplitude linearly scaling as $L/4$ in nm, with and without the tolerable roughness PSD (assumed to be a power-law spectrum $P(f) = K_n/f^n$ with $n = 1.05$ and $K_n = 18$ nm$^3$μm$^{-1.05}$). The component amplitudes have been selected with 1 μrad amplitude/period ratio, and the resulting HEW is always within the 3 arcsec tolerance.

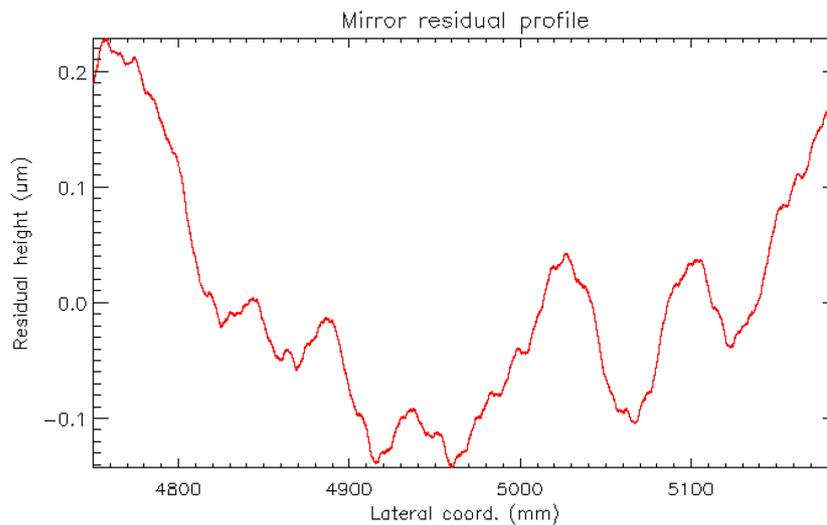

Fig. 8: one of the possible profile errors, obtained adding up 16 harmonics with constant amplitude/period ratio of 1 μrad. This profile error gives us an idea of the tolerable figure error on the longitudinal profile of the mirror.

The paraboloidal mirror (Fig. 9) was purchased from Carl Zeiss SMT GmbH (Oberkochen, Germany), in a preliminary figure (rms < 5 µm) and roughness (rms < 0.5 µm) finishing status. The mirror has recently arrived to our labs, and a metrological analysis (Fig. 10) showed us that the mirror is within the preliminary grinding and lapping specifications. The final, much tighter, polishing level will be reached using the ZEEKO lapping machine installed at INAF/OAB, while the final shape accuracy can be obtained by Ion Beam Figuring using the dedicated facility at OAB.[15] Also a spare grinded mirror - without lapping - was procured from Zeiss to be polished and figured if needed. Initial tests with HOQ 310 samples are in progress at the ZEEKO facility and the first polishing cycle is expected to be started soon.

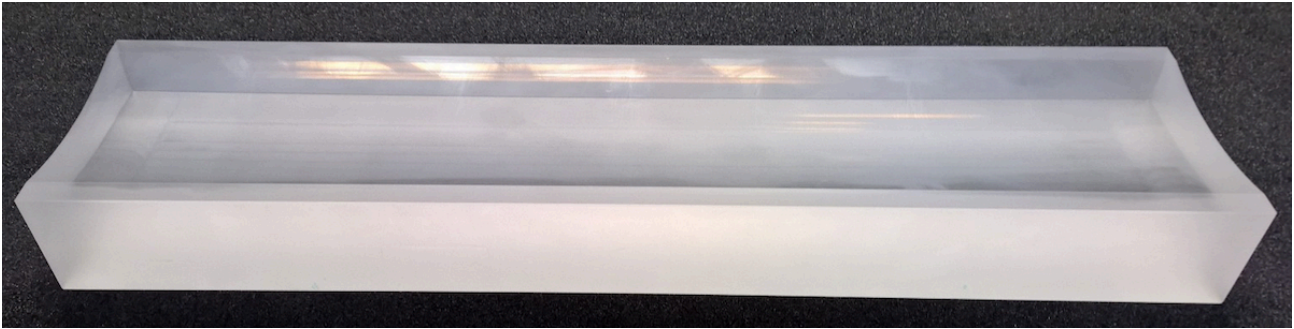

Fig. 9: one of the two paraboloidal mirrors procured from Carl Zeiss SMT GmbH, after the preliminary grinding. The mirror was subsequently lapped to smooth the periodic pattern left after grinding.

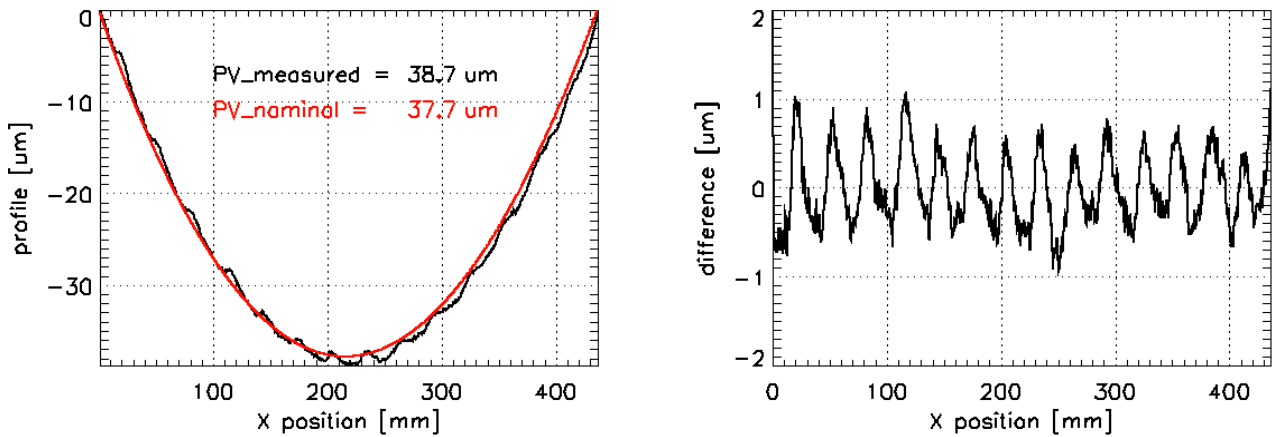

Fig. 10: preliminary metrology on the pre-grinded and pre-lapped mirror as procured from Zeiss. (left) the central measurement obtained with the MPR-700 (Mandrel Profilometer/Rotondimeter)[16] at Media Lario, compared to the nominal longitudinal profile. As per our request, the surface has been delivered in a *preliminary* grinding and lapping state, but already has the correct parabolic shape. (right) the profile error. The high-frequency pattern will be removed by polishing with the ZEEKO machine, while the low-frequency part is expected to be corrected by the IBF facility. Both facilities are already operative at OAB.

### 3.2. The asymmetric cut crystal

The asymmetric cut crystal for the 4.51 keV setup is also being realized. A 20 cm diameter, 5 cm high cylinder in monocrystalline silicon (Fig. 11) has been procured from MEMC. The silicon is extremely pure and the terminal faces of the cylinder correspond to the (100) plane orientation, therefore the (220) planes in the silicon cubic cell are oriented at 45 deg with respect to the terminal surfaces. The beam expander surface has to form a 44.7 deg angle with the (220) planes, therefore the cut plane is just 0.3 deg off the cylinder base. A precise cut is required to ensure the correct expansion factor: the asymmetry angle has to be determined within a 0.1 deg tolerance.

The crystal will be cut using the diamond saw facility operated at CNR-IMEM. The cutting procedure introduces a lattice damage at the crystal surface, which needs to be removed by polishing. The polishing process allows the surface crystal to be smooth enough to be "visibly" shiny but not to a level of smoothness within a few angstroms, as usually requested in grazing incidence optics (Sect. 3.1). Such a smoothness level, on the other hand, is not even required because the diffraction process does not occur at the surface but in the structure of the silicon crystal, that is largely unaffected by residual irregularities of the surface. The final performances of the asymmetric crystal will be directly tested in X-rays at CNR-IMEM.

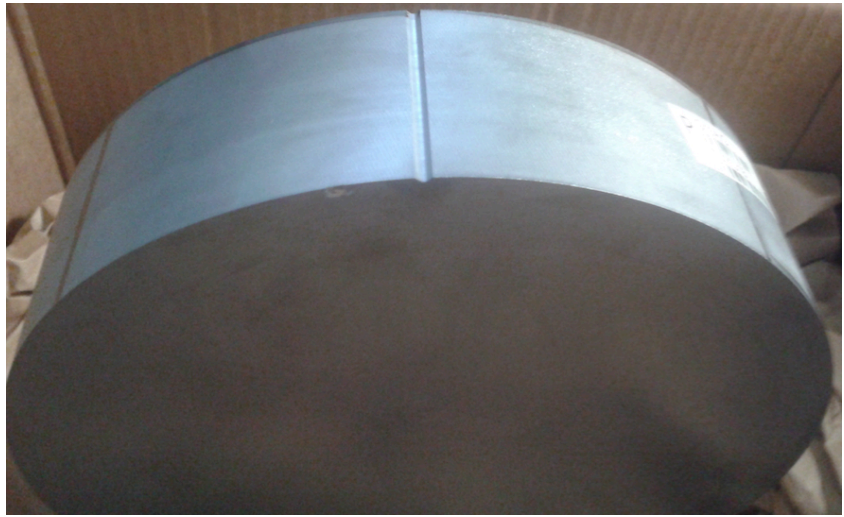

Fig. 11: the monocrystalline cylinder in silicon purchased from MEMC (20 cm diameter. 50 mm thick). The terminal surfaces are the planes (100). The lateral groove locates the (010) planes.

### 3.3. The laboratory

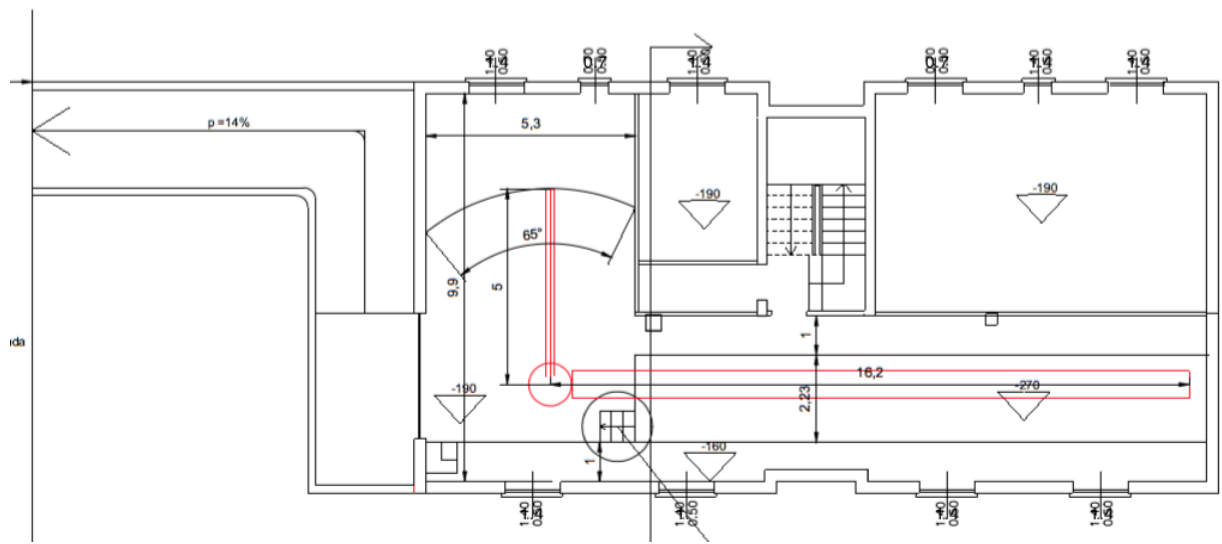

Fig. 12: design of the new laboratory, to be realized on the basement of an existing building. BEaTriX's planned location is drawn in red.

Even if BEaTriX will fit in a small laboratory, this is to be set up properly. The location has been determined as the basement of one of the buildings of INAF-OAB, Merate, formerly housing the INAF-OAB personnel, and currently being turned into laboratories. The underground floor is based on ground and the plans for the renovation and the change of utilization (Fig. 12) are currently complete. The short arm (Sect. 2.2) will be accommodated in one of the existing rooms and, by joining some other rooms, the length of the long arm will conveniently fit the residual space. An outer ramp will also be constructed to pass BEaTriX components (including vacuum tubes) directly from the outer space, once the laboratory is complete.

## 4. SYSTEM SIMULATION

We have simulated the beam conditioning stage of BEaTriX using a code written in IDL language. The code traces rays from the 4.51 keV source in the origin of the reference frame in random directions within the entrance pupil of the parabolic mirror. The reflection off the parabolic mirror is computed according the geometric optics with a survival probability equal to the coating reflectivity. The surviving rays are traced to the symmetric crystals and then to the asymmetric one, reflected in both cases with likelihood depending on their incidence angle, using the reflectivity values shown in Fig. 3 and Fig. 5. The modeling adopted and some traced rays are depicted in Fig. 13.

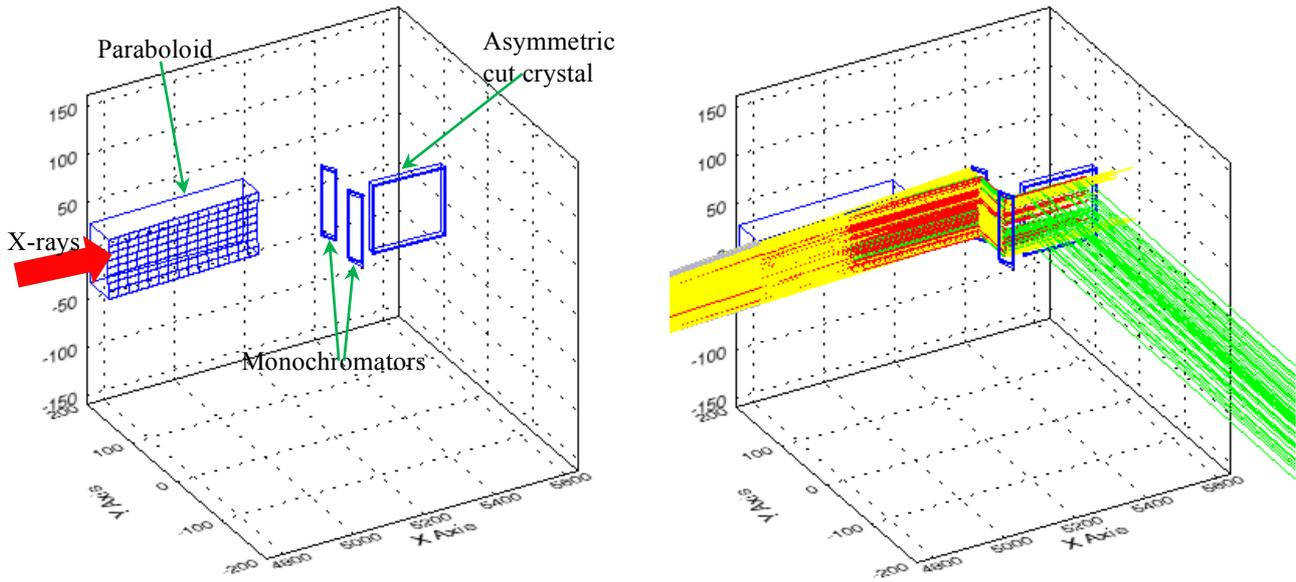

Fig. 13: (left) the current baseline design of BEaTriX as implemented in the IDL simulation code. (right) ray-tracing results: the expanded beam rays are traced in green. Red rays were absorbed in either the residual atmosphere, the collimating mirror, the monochromators, or the asymmetric crystal. Yellow rays have missed some component in the beam conditioning chain and have been discarded.

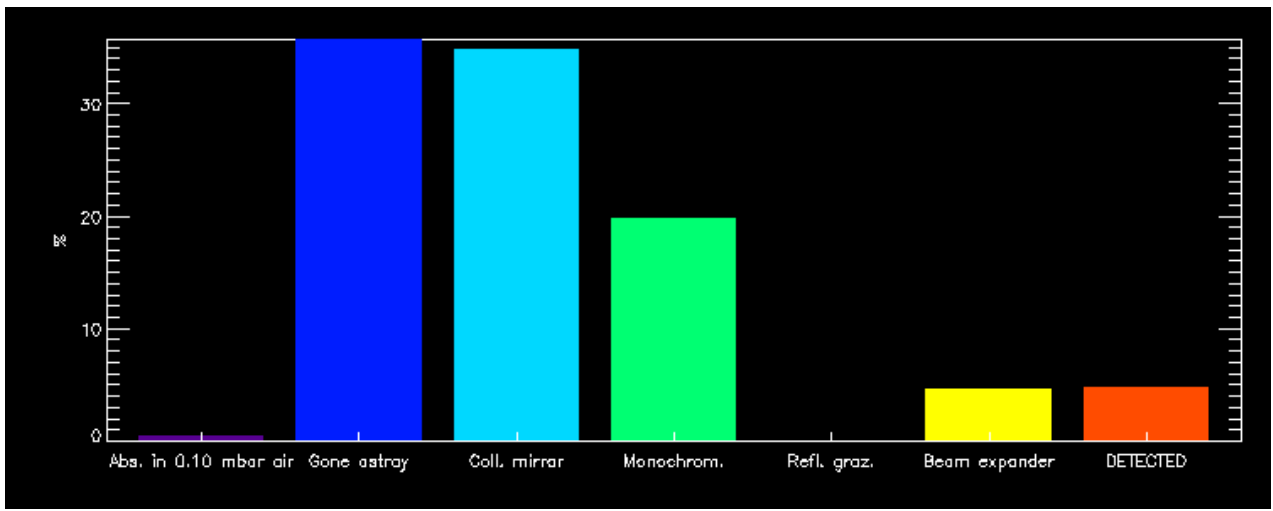

Fig. 14: ray breakdown in the simulated BEaTriX setup. The bar chart shows the percent of X-rays lost in the different stages of the beam conditioning (absorption in the residual atmosphere, lost, absorbed by the mirror, the monochromators, grazing reflected or absorbed by the beam expander). Only the rightmost bar (5%) represent the fraction of rays that have reached the detection plane, at the end of the 12 m long tube.

The simulated source is 30 μm × 30 μm in size, with an isotropic and monochromatic emission at 4.51 keV. The silicon crystals were assumed to be flawless, without misalignments. In contrast, the mirror is assumed to be characterized by a longitudinal figure error returning a 3 arcsec HEW as specified in the fabrication tolerances (Sect. 3.1): as a consequence, the alignment of both monochromators needed to be corrected by 2 arcsec with respect to their nominal orientation. During the propagation, the rays can encounter diverse fates as shown by the chart in Fig. 14: less than 1% is absorbed in the low vacuum (0.1 mbar) assumed in the simulation. A more relevant part (34%) is absorbed by the collimating mirror and approximately the same fraction has missed some optical element. A 20% of traced rays is absorbed in the monochromators, mostly because of the p-component absorption at incidence angles close to the polarization one (45 deg): as a result, the final beam is almost 100% polarized in the vertical plane. A negligible fraction of X-rays undergoes total reflection on the beam expander, but 5% of the traced rays is absorbed in asymmetric diffraction. The remaining 5% finally form the expanded and collimated beam: the performance is higher than in the previous design[9] owing to the higher peak reflectivity of the crystals.

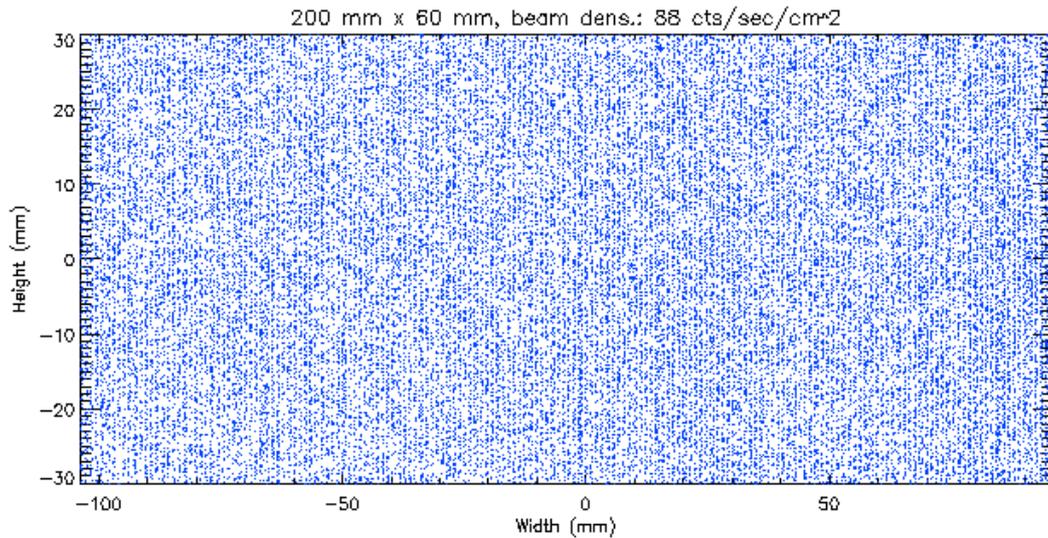

Fig. 15: final positions of the diffracted rays over a plane at a 12 m distance (the focal length of ATHENA) away from the asymmetric crystal. $10^6$ rays were launched, 5% of them have reached the target (Fig. 14). Without a focusing module, the 200 × 60 mm area is covered uniformly: the expected beam density is ~90 ph/sec/cm$^2$ if the source monochromatic radiance is (conservatively) assumed to be $10^{10}$ ph/sec/sterad.

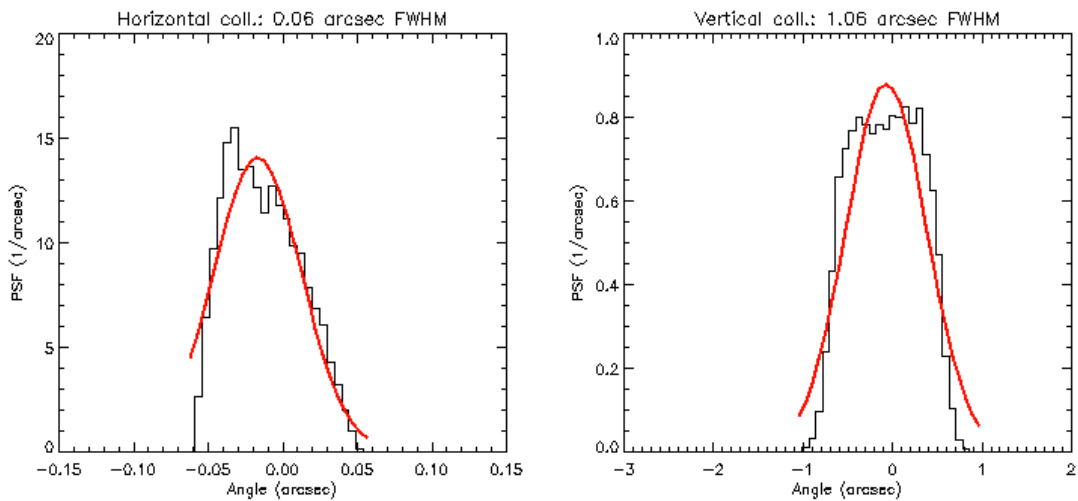

Fig. 16: (left) horizontal and (right) vertical collimation in the finally expanded beam. The graphs represent the distributions of angular deflections with respect to the ideal direction of the collimated beam. The red lines are the respective Gaussian fits. The vertical collimation is dominated by the source size, while the horizontal collimation mostly depends on the asymmetric crystal and the angular spread introduced by the mirror defects.

As a performance metric, we obtain an expanded beam of 200 mm × 60 mm size, quite uniformly covered with collimated rays (Fig. 15). Accounting for the estimated intensity of available X-ray sources (Sect. 2), the achievable beam intensity is 90 to 900 counts/sec/cm$^2$, largely sufficient to provide a fast PSF characterization of an XOU. In fact, even assuming ~6 cm$^2$ as the smallest XOU's effective area and a 90 counts/sec/cm$^2$ beam density imply a ~20 sec integration time to accumulate 10$^4$ counts (the number may vary depending on the detector efficiency), i.e. to achieve a 1% accuracy in the collected photon count.

The simulated beam collimation is very tight (Fig. 16), making it suitable to characterize XOUs with an intrinsic 3-4 arcsec HEW. It should be kept in mind, however, that the mentioned performances refer to the ideal case of a perfectly monochromatic source and a particular profile error on the parabolic mirror; in addition, assuming zero roughness. These factors, if accounted for in the simulation, are expected to degrade the horizontal collimation but, fortunately, only to a minor extent the vertical collimation, which represents the angular spread in the incidence plane of the XOUs under test. Anyway, the mirror tolerances for roughness established in Sect. 3.1 should suffice to keep the beam uniformity and the horizontal collimation within the prescribed limits (Sect. 1). Mirror surface roughness and a polychromatic source will be included in next simulations to determine whether additional monochromators should be added in the BEaTriX setup to shrink the energy bandpass. This will go at the expense of the beam intensity (probably by a 4-fold factor if the number of monochromators is doubled),[10] but since the beam is apparently intense, the corresponding increase in the integration time will be perfectly acceptable. As the surface finish of the collimating mirror progresses, we will be able to include the real surface defects measured by MPR metrology into the simulation and so ascertain that the achieved mirror shape accuracy fulfills the expanded beam specifications.

## 5. CONCLUSIONS

The BEaTriX X-ray facility, being developed at INAF-OAB, will generate a broad, monochromatic (1.49 keV and 4.51 keV), parallel, collimated, and polarized X-ray beam in a small space. The design is being finalized and the realization of the optical components has already started. Once completed, BEaTriX will provide us the capability to perform a direct, non-destructive X-ray characterization of single and stacked mirrors developed for the ATHENA X-ray telescope. The facility will be compact, flexible, and able to operate a characterization of the focusing elements in situ and in real time. Also other applications requiring a broad and parallel beam of soft X-rays can also be envisaged.

## ACKNOWLEDGMENTS

The development of the BEaTriX facility is part of the *AHEAD* consortium activities, financed by the EU Horizon 2020 grant No. 654215. We thank L. Raimondi (ELETTRA/Sincrotrone Trieste) and J. Sutter (Diamond Light Source, UK) for useful discussions.